\documentclass[twocolumn,superscriptaddress,
amsmath,amssymb,
 aps,
prl,]{revtex4-1}
\usepackage{epsfig}
\usepackage{float}
\usepackage{color}
\usepackage{multirow}
\usepackage{amsfonts}
 
\begin{document}

\title{Anomalous symmetry breaking in Weyl semimetal CeAlGe}

\author{H. Hodovanets}
\thanks{Current address: Department of Physics, Missouri University of Science and Technology, Rolla, Missouri 65409 USA}
\affiliation{Maryland Quantum Materials Center, Department of Physics, University of Maryland, College Park, Maryland 20742 USA}
\author{C. J. Eckberg}
\affiliation{Maryland Quantum Materials Center, Department of Physics, University of Maryland, College Park, Maryland 20742 USA}
\author{D. J. Campbell}
\affiliation{Maryland Quantum Materials Center, Department of Physics, University of Maryland, College Park, Maryland 20742 USA}
\author{Y. Eo}
\affiliation{Maryland Quantum Materials Center, Department of Physics, University of Maryland, College Park, Maryland 20742 USA}
\author{P. Y. Zavalij}
\affiliation{X-ray Crystallographic Center, Department of Chemistry and Biochemistry, University of Maryland, College Park, Maryland 2074 USA}
\author{P. Piccoli}
\affiliation{Department of Geology, University of Maryland, College Park, Maryland 20742 USA}
\author{T. Metz}
\affiliation{Maryland Quantum Materials Center, Department of Physics, University of Maryland, College Park, Maryland 20742 USA}
\author{H. Kim}
\thanks{Current address: Department of Physics, Missouri University of Science and Technology, Rolla, Missouri 65409 USA}
\affiliation{Maryland Quantum Materials Center, Department of Physics, University of Maryland, College Park, Maryland 20742 USA}
\author{J. S. Higgins}
\affiliation{Maryland Quantum Materials Center, Department of Physics, University of Maryland, College Park, Maryland 20742 USA}
\author{J. Paglione}
\affiliation{Maryland Quantum Materials Center, Department of Physics, University of Maryland, College Park, Maryland 20742 USA}
\affiliation{Canadian Institute for Advanced Research, Toronto, Ontario M5G 1Z8, Canada}

\begin{abstract}

CeAlGe, a proposed type-II Weyl semimetal, orders antiferromagnetically below 5~K. At 2~K, spin-flop and spin-flip transitions to less than 1~$\mu_B$/Ce are observed in the $M(H)$ data below 30~kOe, \textbf{H}$\|$\textbf{a} and \textbf{b}, and 4.3~kOe,\textbf{H}$\|$ $\langle110\rangle$, respectively, indicating a fourfold symmetry of the $M(H)$ data along the principal directions in the tetragonal \textit{ab} plane with $\langle110\rangle$ set of easy directions. However, anomalously robust and complex twofold symmetry is observed in the angular dependence of resistivity and magnetic torque data in the magnetically ordered state once the field is swept in the \textit{ab} plane. This twofold symmetry is independent of temperature and field hystereses and suggests a magnetic phase transition that separates two different magnetic structures in the \textit{ab} plane. The boundary of this magnetic phase transition and possibly the type of the low-field magnetic structure can be tuned by an Al deficiency.

\end{abstract}

\maketitle

\section{Introduction}

Weyl semimetals have attracted much attention due to their intricate properties associated with the topological manifestation of electronic band structure and their potential application in spintronics, quantum bits, thermoelectric and photovoltaic devices \cite{Wan2011,Weng2015,Hasan2017,Chang2017,Yan2017,Armitage2018}. Magnetic semimetals that break spatial inversion and time-reversal symmetries are relatively scarce, and it is especially hard to confirm the breaking of the time-reversal symmetry in these materials \cite{Wan2011,Witczak2012,Liu2014a,Neupane2014,Wang2016,Manna2018,Liu2018}. The \textit{R}AlGe and \textit{R}AlSi (\textit{R} = Ce, Pr, Nd, and Sm) families present a new class of magnetic Weyl semimetals where both inversion and time-reversal symmetries are broken due to the intrinsic magnetic order \cite{Chang2018,Hodovanets2018,Puphal2019,Yang2020,Lyu2020,Gaudet2021,Wang2022,Zhao_2022,Xu2021a}. With observations of a topological magnetic phase \cite{Puphal2020}, anomalous Hall effect (AHE) \cite{Meng2019}, topological Hall effect \cite{Puphal2020}, and singular angular magnetoresistance (AMR) \cite{Suzuki2019}, with a possible route to axial gauge fields \cite{Destraz2020},  \textit{R}AlGe family is particularly promising.

Noncentrosymmetric CeAlGe, a proposed type-II magnetic Weyl semimetal \cite{Chang2018} that orders antiferromagnetically below 5~K  in zero magnetic field and ferrimagnetically in non-zero field \cite{Hodovanets2018}, hosts several incommensurate multi-\textit{$\vec{k}$} magnetic phases, including a topological phase for \textbf{H}$\|$\textbf{c} \cite{Puphal2020}. 
Motivated by the fact that its magnetic moments lie in the tetragonal \textit{ab}-plane, together with the observation of a sharp singular AMR in its Si-substituted variant \cite{Suzuki2019}, we study pure CeAlGe using magnetization, \textit{M}, angle-dependent magnetic torque, $\tau(\varphi)$, and magnetoresistance, $R(\varphi)$ measurements. While we find the expected fourfold tetragonal symmetry in \textit{M(H)} when field is swept through the \textit{ab} plane, we also observe an anomalous twofold symmetry in both angle-dependent magnetic torque and AMR in the ordered state. In contrast to conventional smoothly changing (i.e. sinusoidal) AMR in magnetic conductors \cite{McGuire1975}, which is dependent on the orientation of magnetization and current, the twofold symmetric \textit{ab}-plane AMR of CeAlGe is remarkably history independent and unchanged under magnetic field and temperature hystereses, highlighting possibilities for device applications. We discuss the idea of the two different magnetic structures in the ordered state as a likely explanation for the observed twofold symmetry, and consider other possibilities.
\section{Experimental details}

\begin{figure}[b]
\centering
\includegraphics[width=1\linewidth]{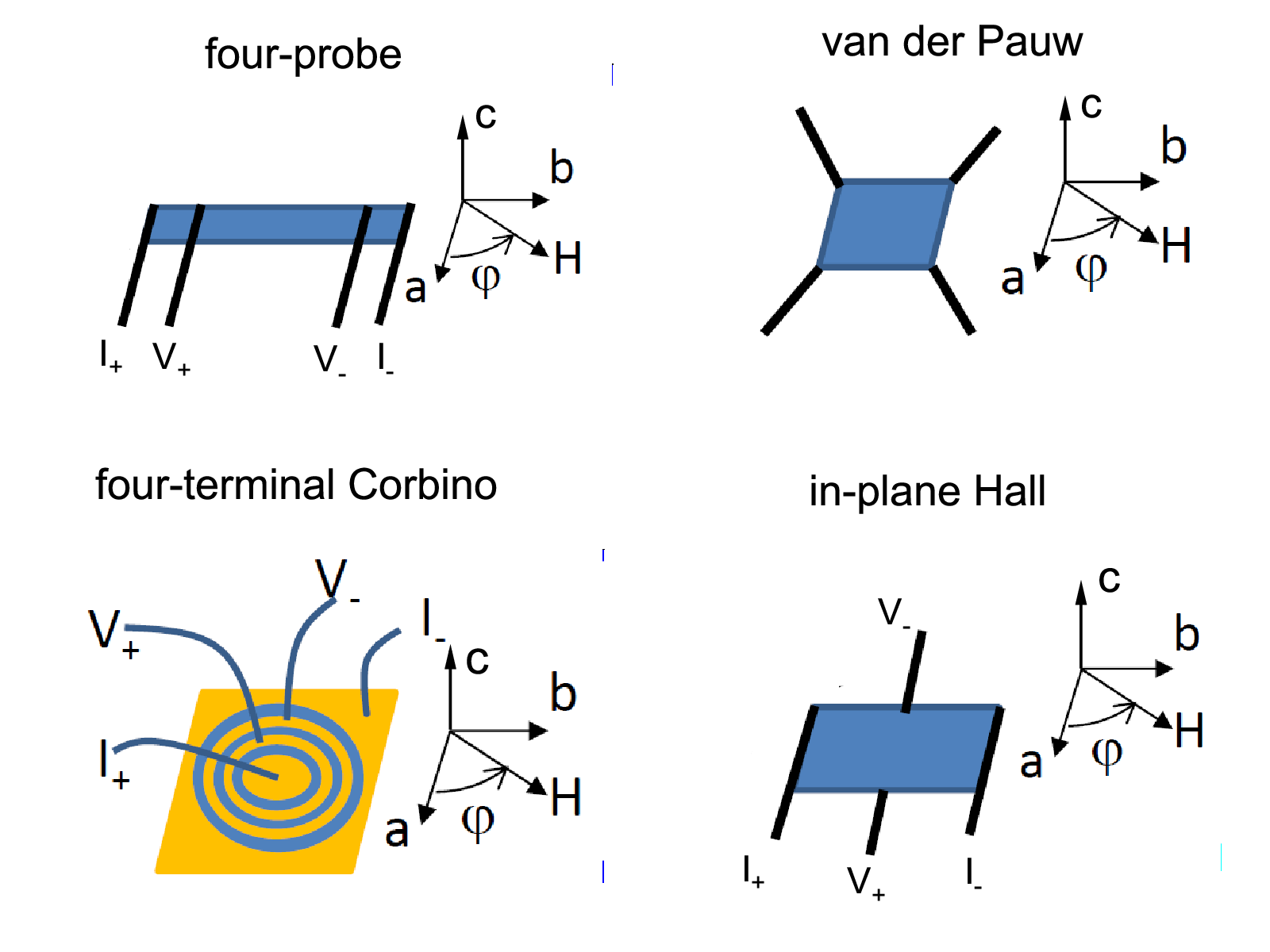}
\caption{\footnotesize (color online) Electrode configurations used in the different resistance as a function of angle measurements.}
\label{DC}
\end{figure}

Single crystals were grown by the high-temperature flux method using Lindberg/Blue M 1500$^{\circ}$ C box furnace as reported in Refs. \cite{Hodovanets2018,Canfield2016}. Temperature-, field-, and angle-dependent magnetization, resistivity, and magnetic torque measurements were performed in a commercial cryostat. All angle-dependent data were collected on changing the angle from 0$^{\circ}$ to 360$^{\circ}$ unless otherwise noted. Resistivity measurements were made in a standard four-probe, van der Pauw \cite{Pauw1958}, Hall bar, or concentric ring (we will call it a four-terminal Corbino) geometry ($I$ = 0.5 or 1~mA), Fig.~\ref{DC}. The samples were polished and shaped with care to not have any Al inclusions. Electrical contacts to the samples in the four-probe, van der Pauw, and the Hall bar geometry were made with Au wires attached to the samples using EPOTEK silver epoxy and subsequently cured at 100$^{\circ}$ C. For the van der Pauw technique \cite{Pauw1958}, the contacts were placed at the corners of the square sample and the current was sent either along the \textit{a} and \textit{b} axes (voltage measured along opposite edges), where the final resistivity was symmetrized, or along a diagonal [110] direction (voltage measured across the opposite diagonal), Hall configuration, where the field was swept in the plane of the sample, and the signal was not symmetrized. For the in-plane Hall configuration sample, the regular Hall bar was prepared, with the electrical current being along the [100] direction; however, the magnetic field was also swept in the plane of the sample instead of being perpendicular to the sample as in the conventional Hall measurement. The four-terminal Corbino was patterned using standard photolithography followed by a standard metal liftoff. The patterns consist of 20-30 \AA/1500 \AA~ Ti/Au contacts made by e-beam evaporation. 25 $\mu$m Au wires were attached to the gold electrodes by wire bonding. To calculate the resistivity of the four-terminal Corbino one needs a geometric factor, which is difficult to estimate when the sample is not a two-dimensional (2D) material or a thin film. To estimate the geometric factor of a single crystal that has a finite thickness, we used the effective thickness that was found numerically \cite{Eo2018}. In the four-terminal Corbino, the current was sent radially in the plane of the sample and the magnetic field was also swept in the plane of the sample as is schematically shown in Fig.~\ref{DC}. The electrical current direction will be specified in the data presented below.

The torque magnetometry option of Quantum Design Physical Properties Measurement System was used to collect magnetic torque data. The sample for this measurement was secured with the help of Apiezon N grease. The background of the Si torque chip with the N grease and sample was measured at 90$^{\circ}$ and was accounted for in the final result.

\begin{figure*}[tbh]
\centering
\includegraphics[width=1\linewidth]{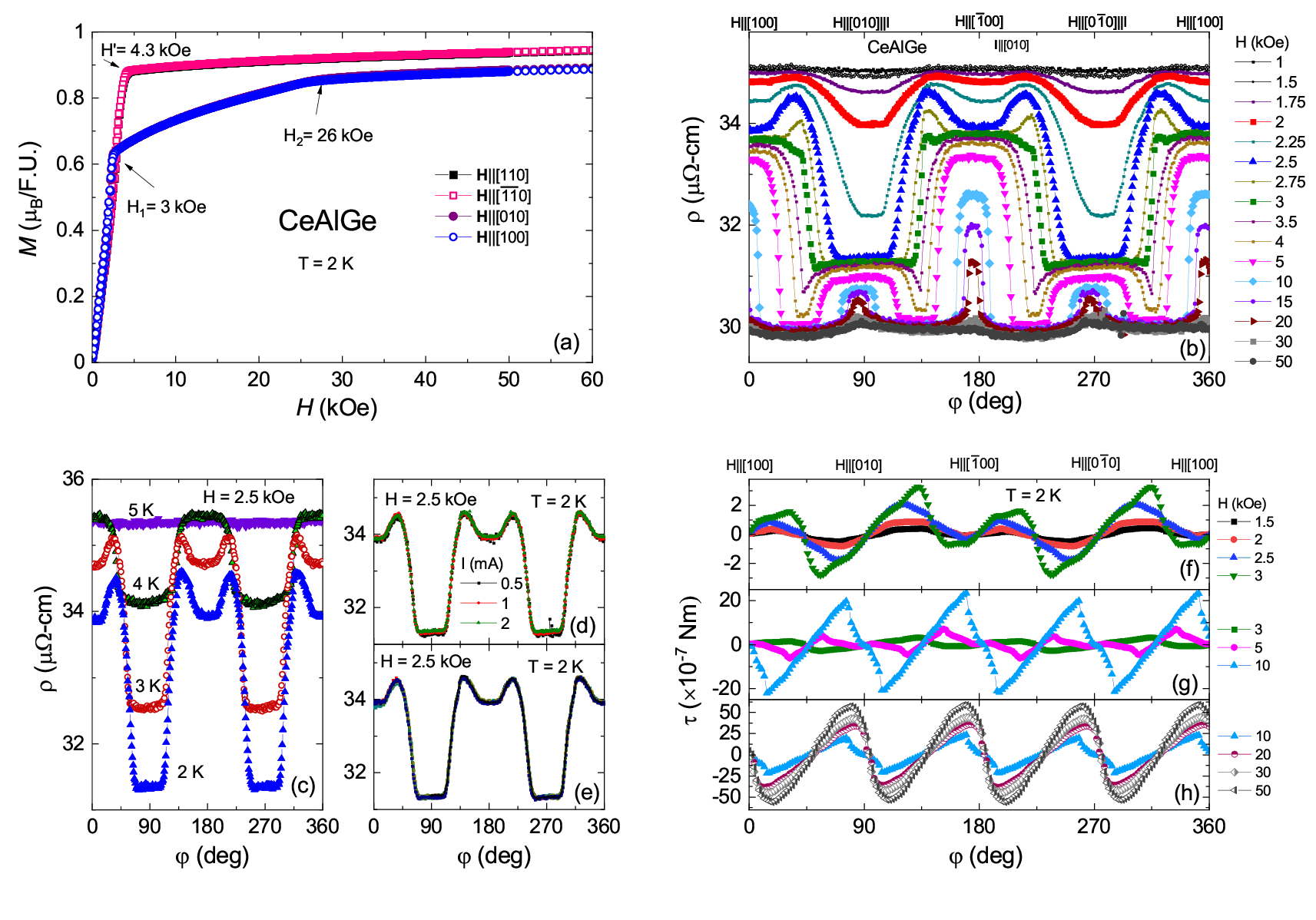}
\caption{\footnotesize (color online) (a) Field-dependent magnetization of CeAlGe for \textbf{H}$\|$\textbf{a}, \textbf{H}$\|$\textbf{b}, \textbf{H}$\|$[110], and \textbf{H}$\|$[$\bar{1}\bar{1}$0] at 2~K. $H_1$ denotes the lowest critical field (beginning of the canted phase of the spin-flop transition) below which the hysteresis in the \textit{M(H)} data starts in the \textbf{H}$\|$\textbf{a} and \textbf{H}$\|$\textbf{b} data. $H_2$ marks a critical field of spin-saturated ferromagnetic state. For \textbf{H}$\|\langle110\rangle$ directions, a spin-flip transition occurs at the critical field marked $H'$. (b) AMR of CeAlGe single crystal measured in the four-probe configuration at \textit{T} = 2~K with \textit{H} = const and \textbf{I}$\|$[010]. The field was swept in the tetragonal \textit{ab} plane. (c) AMR of CeAlGe single crystal measured in the four-probe configuration at \textit{H} = 2.5~kOe and selected \textit{T} = const. AMR of CeAlGe single crystal measured in the four-probe wire configuration at 2~K and 2.5 kOe (d) different current and (e) different conditions on approaching 2 K and 2.5 kOe: cool down from 3~K to 2~K in 2.5~kOe; at 2~K, changed the field from 1.5 to 2.5~kOe; at 2~K, swept through 0 Oe (-30 to 30 kOe); warmed up to 10 K, set \textit{H} = 50~kOe, cooled down to 2 K; warmed up to 10 K, demagnetized with 50 kOe, zero-field cooled to 2 K; warmed up to 10 K at position 90$^{\circ}$, set \textit{H} = 50 kOe, cooled down to 2 K, started off from 0$^{\circ}$; warmed up to 10 K at position 90$^{\circ}$, set H = 50 kOe, cooled down to 2 K, started off from 90$^{\circ}$. (f-h) Angle-dependent magnetic torque data at selected \textit{H} = const collected at 2~K. The data for \textit{H} = 3 and 10~kOe are repeated for clarity.}
\label{MR}
\end{figure*}

\section{Results}
The field-dependent magnetization $M(H)$ data at \textit{T} = 2~K measured for \textbf{H}$\|$\textbf{a}, \textbf{b}, [110], and [$\bar{1}\bar{1}$0] axes are shown in Fig.~\ref{MR}(a). For \textbf{H}$\|$\textbf{a} and \textbf{b} (circles), a clear sharp spin-flop transition to a less than 1~$\mu_B$ saturation moment is observed below $\sim$26~kOe as was reported in Ref. \onlinecite{Hodovanets2018}. The critical fields $H_1$ and $H_2$ delineate the canted moment phase. On the contrary, the spin-flip transition to a slightly higher value of saturated magnetization is observed for \textbf{H}$\|$[110] and [$\bar{1}\bar{1}$0] data (squares) at $H^{'}$ = 4.3~kOe, indicating that the easy axes are the $\langle110\rangle$ set of directions. The data presented in Fig.~\ref{MR}(a) indicate a four-fold symmetry of \textit{M(H)} data in the tetragonal \textit{ab} plane.

Contrary to $M(H)$ data, a sharp twofold symmetry change is observed in the AMR data when the magnetic field is swept in the tetragonal \textit{ab} plane, Fig.~\ref{MR}(b). This twofold symmetry sets in before the critical fields of $H_1$ and $H^{'}$ (defined from the magnetization data) and is less apparent above the critical field  $H_2$ where the moments are in the field-saturated ferromagnetic state. Keeping the magnetic field constant at \textit{H} = 2.5~kOe, the AMR was measured at constant temperatures, Fig.~\ref{MR}(c). The twofold symmetry holds only in the ordered state, below 5 K, thus suggesting that the origin of this behavior is due to the magnetic order. Neither the magnitude of the current nor the different conditions at which the 2 K temperature is reached nor at what angle the measurement is started have an effect on the AMR features, at least for the \textit{H} = 2.5 kOe data as shown in Figs.~\ref{MR}(d) and (e), respectively.

\begin{figure}[tb]
\centering
\includegraphics[width=1\linewidth]{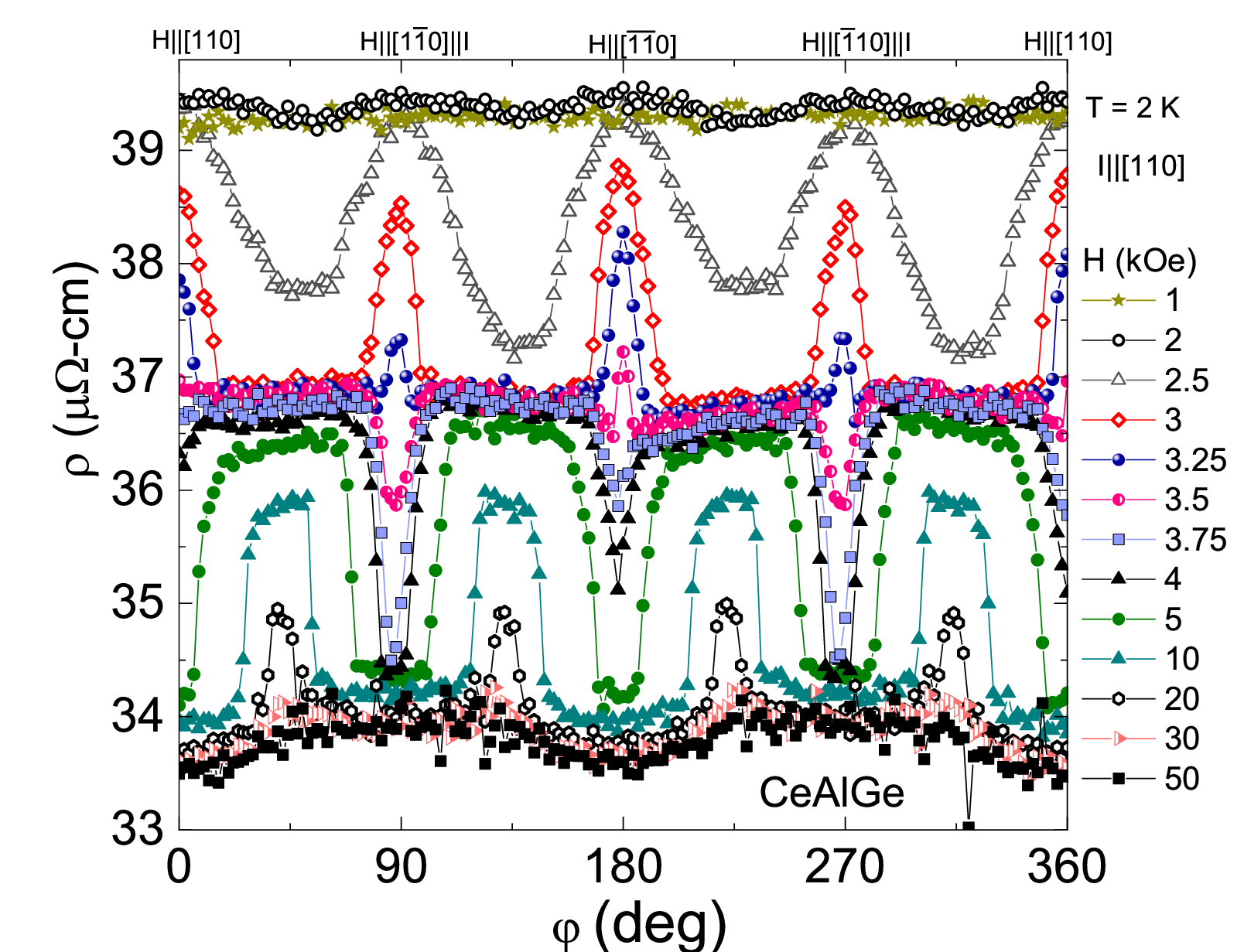}
\caption{\footnotesize (color online) AMR of CeAlGe single crystal measured in the four-probe wire configuration at 2~K, \textbf{I}$\|$[110]. The AMR data are clearly split into two regimes at \textit{H} = 3.5 kOe - a value of the phase transition between the two different magnetic states. Field-dependent resistivity of CeAlGe measured in the four-probe wire configuration at constant angles at 2 K.}
\label{RH}
\end{figure}

To further validate whether the twofold symmetry in the AMR is due to the magnetic order or due to the current direction, we measured magnetic torque in the tetragonal \textit{ab} plane at 2~K with \textit{H}=const, Figs.~\ref{MR}(f)-(h). Figure~\ref{MR}(f) shows $\tau(\varphi)$ data below $H\leq$~3~kOe with a clear twofold symmetry and very complicated functional dependence that cannot be fit by a series of even sine functions\cite{Okazaki2011,Kasahara2012}. The magnetic torque changes the location of positive and negative maxima (sign change) between 3 and 5 kOe, Fig.~\ref{MR}(g). This corresponds in the AMR to the appearance of dips that turn into plateaus at 45$^{\circ}$ (every 45$^{\circ}$) in Figs.~\ref{MR}(b) between \textit{H} = 3.5 and 4 kOe. This region separates the data into two different magnetic regimes and is more evident in the AMR of the sample for which \textbf{I}$\|$[110], Fig.~\ref{RH}. Here, for \textit{H} = 3.25 kOe, all peaks at every 90$^{\circ}$ point up and the same peaks point down at \textit{H} = 3.75 kOe. However in between, at \textit{H} = 3.5 kOe, they alternate, the ones at 90$^{\circ}$ and 270$^{\circ}$ turn down but the ones at 180$^{\circ}$ and 0$^{\circ}$/360$^{\circ}$ stay up. Perhaps the narrow region between $H_1$ and $H'$ critical fields in the \textit{M(H)} data of spin reorientation is captured here. This phase transition at \textit{H} = 3.5 kOe seems to divide AMR into two different regimes. This value of magnetic field is slightly above \textit{H$_1$}(\textbf{H}$\|$\textbf{a}) and much lower than \textit{H$'$}(\textbf{H}$\|$[110]) in the \textit{M(H)} data, Fig.~\ref{MR}(a). The AMR for CeAlGe cannot be simply scaled based on the field-induced magnetization along either the \textit{a} axis or [110] axis (see Fig.~\ref{RH1}).

The twofold anisotropy in the torque data decreases and is barely observable at 10 kOe, Fig.~ \ref{MR}(g), as the magnitude of the torque increases with the magnetic field. As opposed to the AMR, the magnetic torque data display a clear fourfold symmetry at a lower magnetic field \textit{H} = 20 kOe and above, Fig.~ \ref{MR}(h). These observations point to the magnetic order being a culprit of the breaking of the fourfold symmetry in the measurements (according to neutron studies \cite{Suzuki2019,Puphal2020}, the crystal structure of CeAlGe remains tetragonal down to 2 K). This twofold symmetry is more dramatic and enhanced in the resistivity measurements.

\begin{figure}[b]
\centering
\includegraphics[width=1\linewidth]{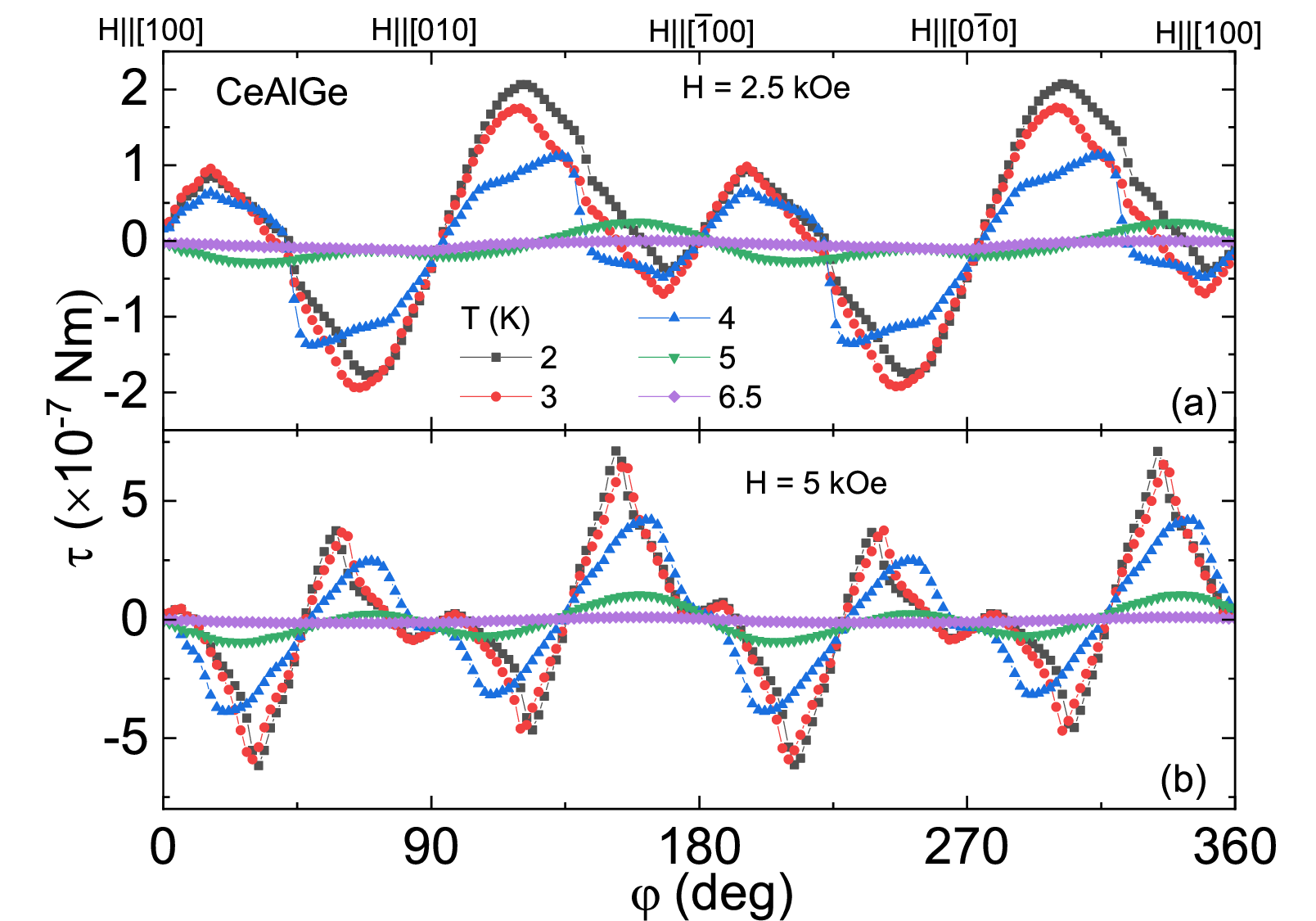}
\caption{\footnotesize (color online) Angle-dependent magnetic torque of CeAlGe single crystal measured at (a) \textit{H} = 2~kOe and (b) \textit{H} = 5 kOe.}
\label{T}
\end{figure}

A markedly different evolution, reflecting different magnetic state, of the angle-dependent magnetic torque with temperature at \textit{H} = 2.5 and 5 kOe is shown in Figs.~\ref{T}(a) and \ref{T}(b), respectively. The absolute value of the torque decreases as the temperature increases. At 5 K, which is just above the ordering temperature, the torque data for these two magnetic fields become similar and at 6.5 K, the torque is almost zero reflecting the paramagnetic state. Interestingly, the torque data at 5 K is still twofold symmetric, perhaps indicating some magnetic moment fluctuations.

\begin{figure}[tbh]
\centering
\includegraphics[width=1\linewidth]{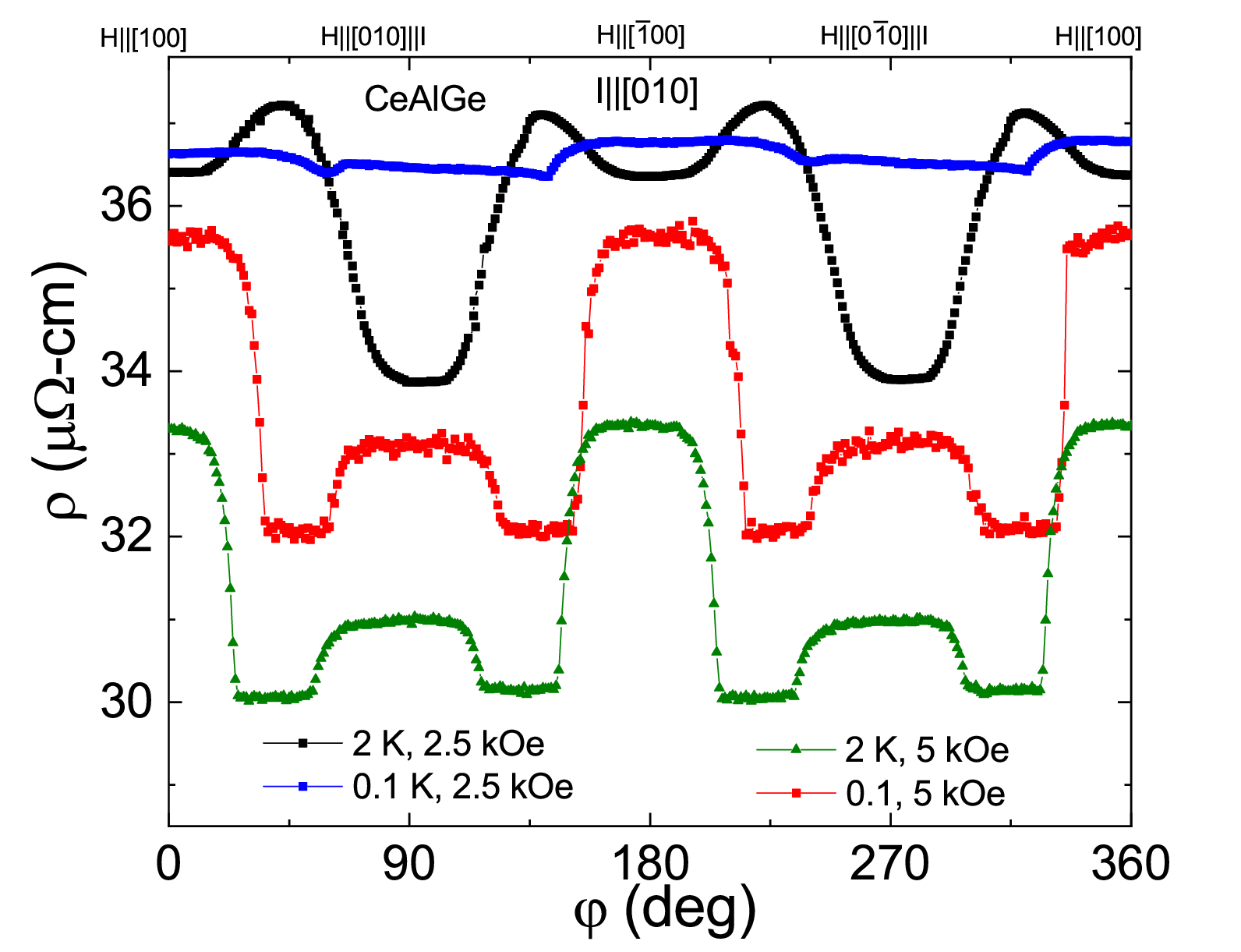}
\caption{\footnotesize (color online) Angle-dependent resistivity of CeAlGe single crystal measured in four-probe wire configuration, \textbf{I}$\|$[010], at $T\leq$2~K for \textit{H} = 2.5 and 5 kOe.}
\label{RHD}
\end{figure}

Upon lowering the temperature to 0.1 K, the shape of the AMR for \textbf{I}$\|$\textbf{b} changes at \textit{H} = 2.5 kOe, Fig.~\ref{RHD}. However, the shape of AMR at \textit{H} = 5 kOe remains unchanged except the peaks become narrower. This indicates another magnetic phase below 2 K with the critical field less than 2.5 kOe.

\begin{figure*}[tbh]
\centering
\includegraphics[width=1\linewidth]{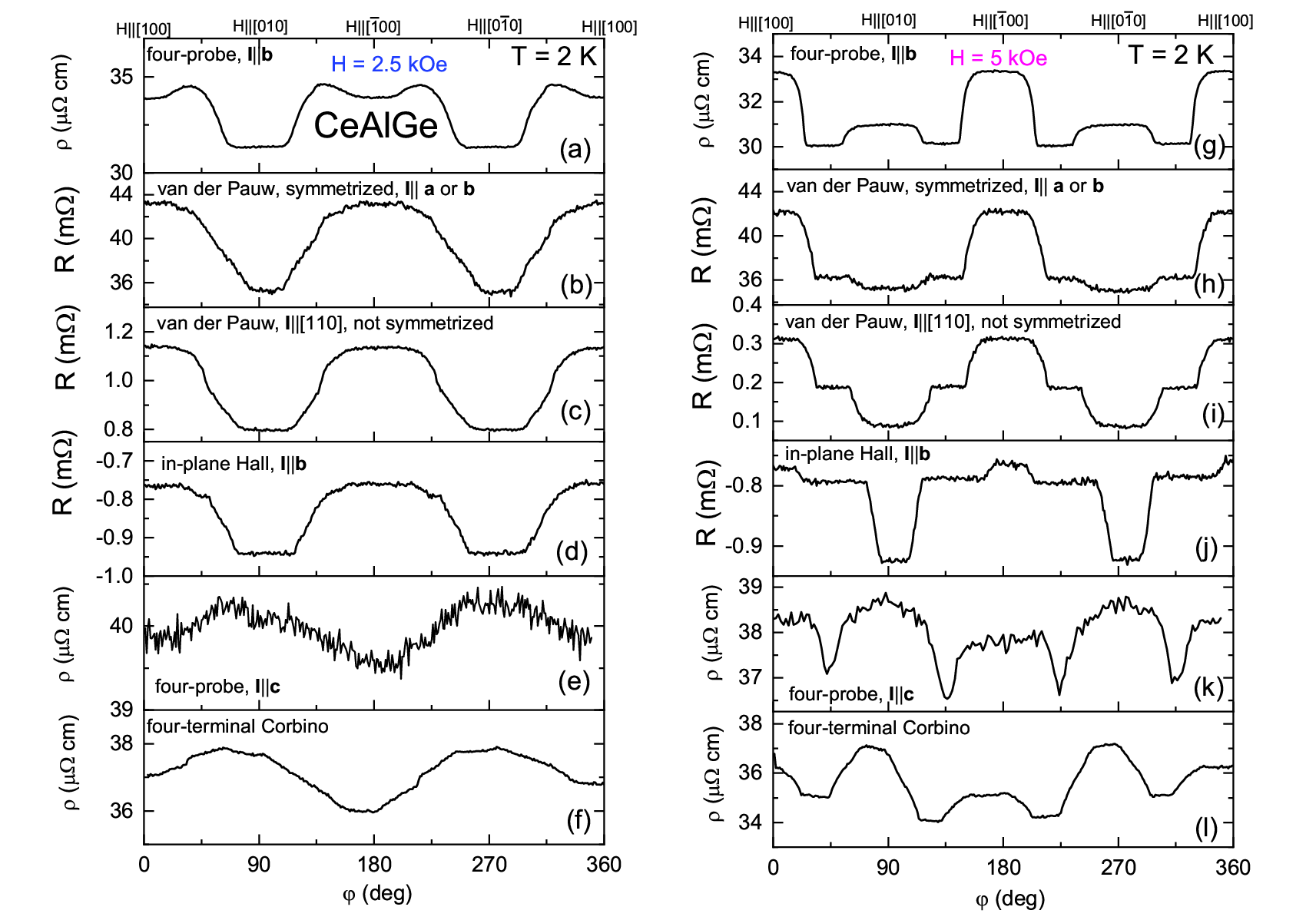}
\caption{\footnotesize (color online) The AMR for various electrode configurations and measurement techniques of CeAlGe single crystals measured at \textit{T} = 2~K with (a)-(f) \textit{H} = 2.5~kOe and (g)-(l) \textit{H} = 5~kOe. The data for the van der Pauw configuration with \textbf{I}$\|$\textbf{a}, \textbf{b}, and [110] were measured on the same sample. The sample for the four-terminal Corbino measurement was mounted, due to the size restrictions, with approximately -13$^{\circ}$ offset with respect to the \textit{a} axis. All samples are from the same batch.}
\label{DM}
\end{figure*}

To further test the effect of the current and its direction, we measured the AMR with different electrode configurations and techniques. The results are shown in Fig.~\ref{DM} for \textit{H} = 2.5 and 5 kOe left and right panels, respectively. The twofold symmetry is present in all measurements. The shape of the resistance as a function of the angle is similar for measurements where the current is send along the \textit{a}-, \textit{b}-, and [110] axes, Figs.~\ref{DM}(a-d), at $H$ = 2.5 kOe with that of four-probe, \textbf{I}$\|$\textbf{b}, displaying more features. However, the position of maxima and minima in the data are shifted for the four-probe,\textbf{I}$\|$\textbf{c}, and four-terminal Corbino samples. At this field, it appears that if the current is sent along the well defined primary in-plane direction, the shape of the AMR will be more or less the same. However, sending the current radially in-plane and along the \textit{c} axis, averages the data and a smooth sinusoidal dependence is observed.

The data for $H$ = 5 kOe, Figs.~\ref{DM}(g)-\ref{DM}(l) don't have such clear separation with respect to the current direction. Here, the data for the four-probe, \textbf{I}$\|$\textbf{b} and \textbf{I}$\|$\textbf{c}, and four-terminal Corbino samples show a similar behavior where the plateaus every 90$^{\circ}$ are concave down, although the height difference is about 90$^{\circ}$ off-phase for the four-probe, \textbf{I}$\|$\textbf{b}, sample.

The other three samples, the data for which is shown in Figs.~\ref{DM}(h)-\ref{DM}(j) show also a similar functional behavior but the height of the peaks varies. For the van der Pauw sample, \textbf{I}$\|$\textbf[110], and the in-plane Hall sample, the negative minima at 90$^{\circ}$ and 270$^{\circ}$ are more pronounced compared to those for the van der Pauw sample for which \textbf{I}$\|$\textbf{a} or \textbf{I}$\|$\textbf{b}. Since for 90$^{\circ}$ and 270$^{\circ}$ angles the magnetic field is parallel to the current, the voltage contacts on the opposite corners of the samples for the former two samples are measuring the potential drop across all sample (across all current lines) and the voltage contacts of the latter sample (the van der Pauw sample for which \textbf{I}$\|$\textbf{a} or \textbf{I}$\|$\textbf{b}) measure only a fraction of the current lines because the voltage contacts are placed on the same side of the sample and are parallel to the current lines/current contacts. When the current and magnetic field are perpendicular, 0$^{\circ}$ and 180$^{\circ}$, the in-plane Hall contribution is small; however, the voltage contacts for the two van der Pauw samples are also measuring longitudinal magnetoresitance, giving a large contribution to the magnitude of the peaks. The AMR is flat in between the maxima and minima. Such behaviour of the AMR is possibly related to the different magnetic order or spin state compared to that at $H$ = 2.5 kOe and it may reflect the fact that the moments are locked in some directions.

\section{Discussion}
Neutron diffraction experiments have reported differing results on the magnetic structure of CeAlGe.
While Ref.~\onlinecite{Suzuki2019} reported a zero-field coplanar $Fd'd2'$ magnetic structure (i.e. \textbf{m} = ($m_x,m_y$,0) on sublattice A and (-$m_x,-m_y$,0) on sublattice B related by a diamond glide-operation) and a collinear magnetic structure (\textbf{m}$\|$[100]) in non-zero field with independent moments on the symmetrically nonequivalent A and B sublattice sites, Ref.~\onlinecite{Puphal2020} reported an incommensurate multi-$\vec{k}$ magnetic ground state in zero field. This magnetic phase changes to a single-$\vec{k}$ state at the metamagnetic transition $H_1$ = 3 kOe (\textbf{H}$\|$\textbf{a}), consistent with our results. The single-$\vec{k}$ state evolves into the field polarized ferromagnetic state at $H_2\sim$9 kOe (lower than $H_2$ for this work) at 2 K. Thus, the two different regimes seen in the AMR data would correspond to these two different magnetic phases. Note that the sample studied in Ref.~\onlinecite{Puphal2020} is stoichiometric and the ones in this work have $\sim 5 \%$ deficiency in both Al and Ge (see Table II in the Appendix). Despite slightly different magnetic ordering temperatures reported in Ref.~\onlinecite{Puphal2020} and in our work here, the critical field $H_1$ is the same. As we discussed above, the field close to $H_1$ determines the boundary of the two magnetic phases in the tetragonal \textit{ab} plane. As is discussed in the Appendix, a large Al deficiency, which depends on different crystal growth conditions, changes the value of $H_1$ and $H'$, making them smaller, and perhaps changing the values of multi-$\vec{k}$ vectors (or magnetic structure altogether) since the features in the AMR in the lower-field state become slightly different. The magnetic phase above these two fields remains unchanged. Systematic magnetic structure studies are needed to confirm this hypothesis. 
 
In Ref.~\onlinecite{Suzuki2019}, the observed singular AMR was suggested to arise, under particular conditions, from momentum space mismatch across real space domains and was confined to very narrow angles. These domains form a single domain once the field is increased so that the sample is in the field saturated ferromagnetic state. One may assume that if the magnetic field is subsequently lowered, the sample would break into a different set of magnetic domains and hence upon remeasuring the AMR, a different functional dependence would be observed. Instead, we still observe the same behavior no matter how many times the sample is warmed up above the ordered temperature and at which field the sample is cooled to 2 K, Fig.~\ref{MR}(e). It is plausible that structural defects (e.g. micro-cracks in the samples after polishing since the samples are rather brittle) or some arrangements of sub-micron Al inclusions may constrict the formation of the domains and once domains are formed, they are pinned and could only be changed if the defects are removed, e.g. by controlled annealing in the former case. Such studies, together with the visualization of magnetic domains \cite{Yang2020a} and defects or pinning centers in the ordered state at constant applied magnetic field, would be necessary. Alternatively, single crystals of CeAlGe can be grown using a different flux (we discuss In-flux grown single crystal of CeAlGe in the Appendix) or a different single-crystal growth technique can be utilized.

Somewhat similar AMR was reported for the magnetic topological semimetal CeBi \cite{Lyu2019}. However, for CeBi, the angle of the negative magnetoresistance is solely dictated by the field-induced magnetization along the \textit{a} axis that flops under the influence of the rotating magnetic field and thus can be scaled.\cite{kuthanazhi2019}. The AMR for CeAlGe cannot be simply scaled based on the field-induced magnetization along either the \textit{a} or [110] axis (see Fig.~\ref{RH1}).

Given that the critical field between the two magnetic structures in the \textit{ab} plane decreases as the temperature is increasing, Fig.~\ref{110}, and is following the line between the two magnetic phases in Fig. 6(a)\cite{Hodovanets2018}, the origin of the observed behavior in the AMR is due to different magnetic orders/spin structure. The enhancement of the magnetoresistane near the spin-flip transition for CeAlGe can be perhaps described theoretically similarly to those of EuCo$_2$As$_2$ and Ca$_{0.9}$Sr$_{0.1}$Co$_2$As$_2$ which also show anomalous features in the AMR due to the magnetic order \cite{Kim2022,Oh2022} pointing to a universality of such occurrence. The sharp step-like change in the AMR and its tunability makes CeAlGe promising for spintronics application.

\section{Conclusion}
A clear twofold symmetry is observed in the robust and sharp nonsinusoidal AMR data in the ordered state when the magnetic field is swept in the tetragonal \textit{ab} plane, revealing  more detailed and complicated underlying magnetic structures and the phase transitions between them. The current along the \textit{b} axis enhances this twofold symmetry compared to the current along the [110] and \textit{c} axes, although the [110] axis is an easy axis. A clear separation of the AMR data into two regimes based on the two distinct magnetic phases is observed in the magnetic torque and the AMR data at the magnetic field close to $H_1$ and $H'$. Al deficiency controls the value of these two critical fields and changes the critical field of the phase transition between the two different magnetic phases and most likely alters the magnetic structure of the low-field phase in the proposed type-II Weyl semimetal CeAlGe.

\section{Acknowledgments} 

The authors would like to thank P. P. Orth, J. Checkelsky, T. Voita, and J. Lynn for insightful discussions. H.H. would like to thank D. M. Benson and B. L. Straughn for fruitful discussions. Materials synthesis efforts were supported by the Gordon and Betty Moore Foundation's EPiQS Initiative through grant no. GBMF9071, and experimental investigations were supported by the Department of Energy, Office of Basic Energy Sciences, under Award No. DE-SC0019154.
\appendix
\newpage

\section{Appendix}
   
\section{Angle- and field-dependent resistivity}

\begin{figure}[tb]
\centering
\includegraphics[width=1\linewidth]{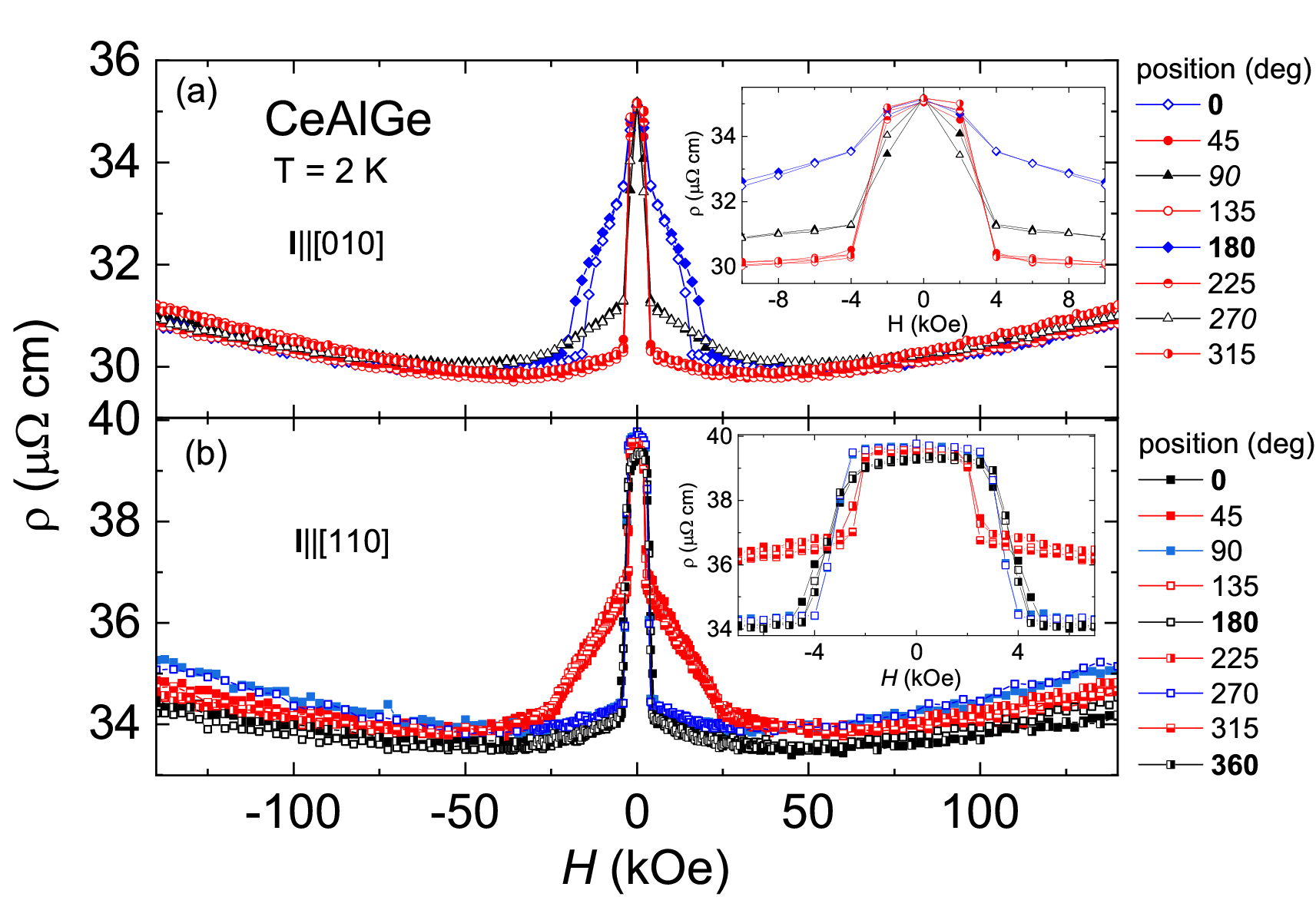}
\caption{\footnotesize (color online)  Field-dependent resistivity of CeAlGe single crystals measured in the four-probe wire configuration at constant angles at 2 K (a) \textbf{I}$\|$[010] and (b) \textbf{I}$\|$[110]. Insets show zoom in of low-field data.}
\label{RH1}
\end{figure}

The field-dependent resistivity data collected every 45$^{\circ}$ at 2 K are shown in Figs.~\ref{RH1}(a) \textbf{I}$\|$\textbf{b} and \ref{RH1}(b) \textbf{I}$\|$[110], respectively. The data fall into three manifolds for \textbf{I}$\|$\textbf{b} and two manifolds for \textbf{I}$\|$[110] in the magnetically ordered state (below field saturated ferromagnetic state). For \textbf{I}$\|$\textbf{b}, Fig.~\ref{RH1}(a), as opposed to the \textit{M(H)} data, $\rho(H)$ data show a clear twofold symmetry for \textbf{H}$\|$[100] and \textbf{H}$\|$[010] directions and a clear four-fold symmetry for \textbf{H}$\|\langle110\rangle$ directions in the ordered state. This would be consistent with the current breaking the fourfold symmetry for 90$^{\circ}$ rotations. On the contrary, for \textbf{I}$\|$[110], Fig.~\ref{RH1}(b), the $\rho(H)$ data seem to follow \textit{M(H)} behavior except below \textit{H} = 3 kOe, inset to Fig.~\ref{RH1}(b), where the data for 0$^{\circ}$ (180$^{\circ}$ and 360$^{\circ}$) and 90$^{\circ}$(270$^{\circ}$) are not the same, i.e. the fourfold symmetry is broken. The difference in the data is not due to the hysteresis since the data were collected with the same approach. Thus, the twofold symmetry in the $\rho(H)$ data for \textbf{I}$\|$[110] is more subtle. Negative magnetoresistance in the ordered state is followed by a positive magnetoresistance at the field close to $H_2$ with almost no anisotropy in the field-saturated state for \textbf{I}$\|$\textbf{b} and small anisotropy for \textbf{I}$\|$[110]. The features in the ordered state are consistent with those observed in the \textit{M(H)} data shown in Fig.~2(a), except for \textbf{I}$\|$\textbf{b} sample at 0$^{\circ}$ and 180$^{\circ}$ a clear sharp change in the AMR is seen at about 16 kOe. There is no corresponding sharp feature in the \textit{M(H)} data Fig.~2(a). The hysteresis in the data below \textit{H} = 4 kOe, consistent with that seen in the \textit{M(H)} data, is evident in the data shown in the inset to Fig. ~\ref{RH}(a). One thus should expect hysteresis on increasing and decreasing the angle in the AMR data as well.

\begin{figure}[t]
\centering
\includegraphics[width=1\linewidth]{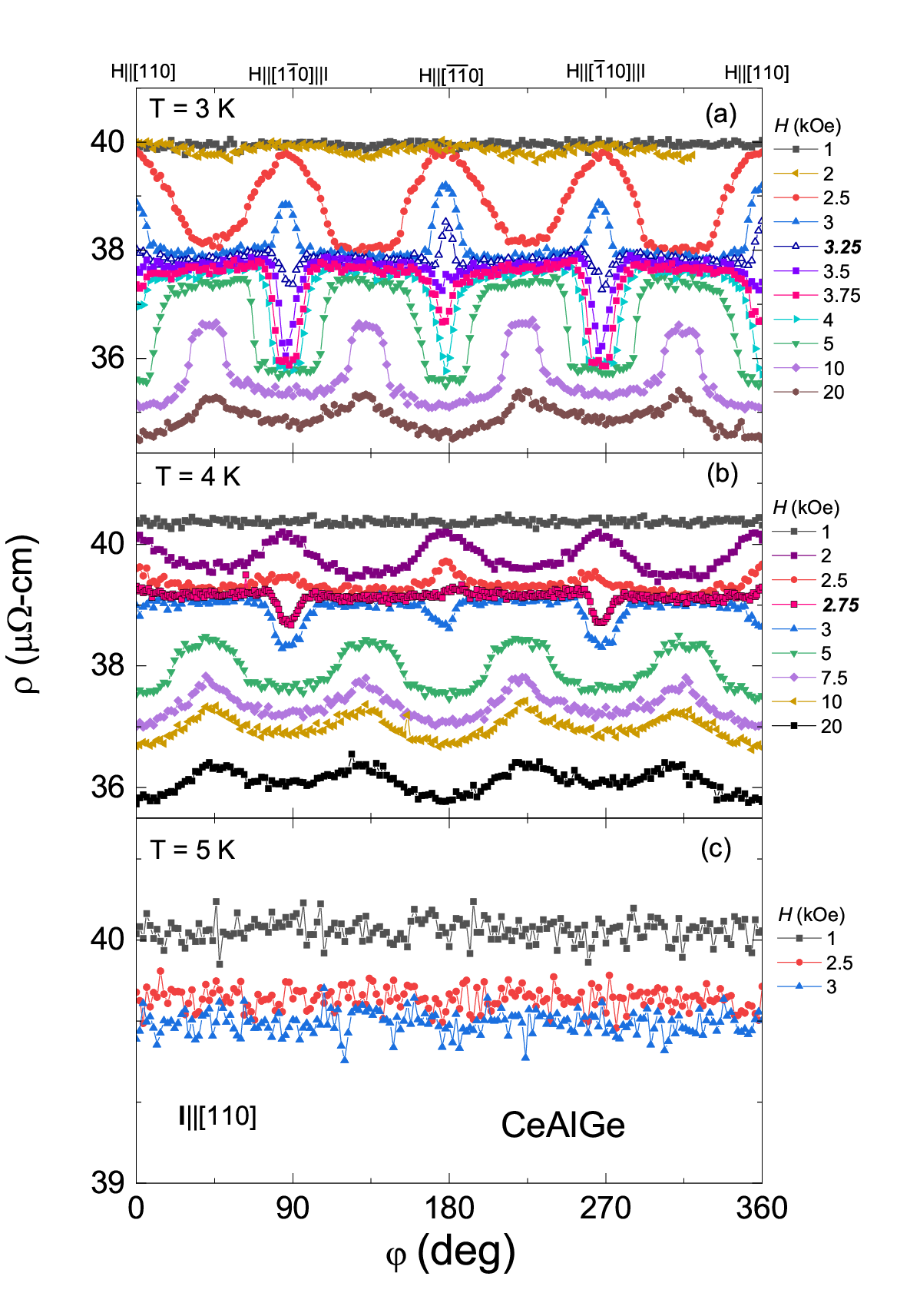}
\caption{\footnotesize (color online) AMR of CeAlGe measured in the four-probe wire configuration, \textbf{I}$\|$[010], with \textit{H} = const at (a) 3 K, (b) 4 K and (c) 5 K.}
\label{110}
\end{figure}

The AMR collected at constant magnetic fields and temperatures for the sample in the four-probe wire configuration with \textbf{I}$\|$[110] is shown in Fig.~\ref{110}. The magnetic fields at which the transition between the two magnetic phases in the tetragonal \textit{ab}-plane occurs is marked in bold font. There is no change in the resistivity as a function of angle when the temperature is raised to 5 K, just above the ordering temperature.

\section{Single crystals grown under different conditions}

Additional two batches of CeAlGe single crystals were grown with the following conditions: (i) cerium ingot from Ames Laboratory and Al flux and (ii) cerium ingot from Alfa Aesar and In flux were used to grow CeAlGe single crystals. Cerium from Alfa Aesar was also used to grow the crystals presented in the main text. A Canfield crucible set with the frit\cite{Canfield2016} was used in these two growths to prevent Si substitution from the quartz wool in the catch crucible. The batches of crystals were grown at different times with the same temperature profile in the same Lindberg/Blue M 1500$^{\circ}$C box furnace. 

Lattice parameters determined through single crystal x-ray diffraction analysis are shown in Table~\ref{tabl1}. A single crystal with Ames Ce shows an only slightly larger \textit{c} axis. In-flux grown sample has both lattice parameters smaller than Al-flux grown samples. 

\begin{table}[t]
\caption{\label{tabl1} Lattice parameters data determined through single-crystal x-ray diffraction of CeAlGe single crystals grown with different conditions. Space group I4$_1$md (No. 109). All data were collected at 250~K on Bruker APEX-II CCD system equipped with a graphite monochromator and a MoK$\alpha$ sealed tube (wavelength $\lambda$~=~0.71070~$\mathrm{\AA}$).}
\footnotesize\rm
\begin{ruledtabular}
\begin{tabular}{llll}
Lattice parameters&Frit (main text)&Ames Cerium/frit&In flux/frit\\
\hline
$a$($\mathrm{\AA}$)&4.2920(2)&4.2930(2)&4.2875(2)\\
$b$($\mathrm{\AA}$)&4.2920(2)&4.2930(2)&4.2875(2)\\
$c$($\mathrm{\AA}$)&14.7496(4)&14.7631(7)&14.7197(7) \\
\end{tabular}
\end{ruledtabular}
\end{table}

\begin{table}[t]
\caption{\label{tabl2} Results of WDS with two standard deviations for CeAlGe single crystals grown under different conditions. The rows represent different samples.}
\footnotesize\rm
\begin{ruledtabular}
\begin{tabular}{llll}
Chemical elements&Frit (main text) &Ames Ce/frit&In flux/frit\\
\hline
Ce&1&1&1\\
Ge&0.95(1)&0.96(2)&0.98(1)\\
Al&0.98(2)&0.83(2)&0.90(1) \\\hline
Ce&1&1&\\
Ge&0.95(4)&0.95(1)&\\
Al&0.95(1)&0.81(2)& \\
\end{tabular}
\end{ruledtabular}
\end{table}

To determine the stoichiometry of the samples, single crystals were analyzed using a JEOL 8900R electron probe microanalyzer at the Advanced Imaging and Microscopy Laboratory (AIMLab) in the Maryland Nanocenter using standard wavelength dispersive spectroscopy (WDS) techniques. The following analytical conditions were utilized: 15 kV accelerating voltage; 50 nA sample current; a 1 micron beam; and synthetic Al and Ge metal and CePO$_4$ standards. Both K-alpha (Al, Ge) and L-alpha (Ce) x-ray lines were used. Count times ranged from 20-30 s on peak, and 5-10 s on the background. Raw x-ray intensities were corrected using a standard ZAF algorithm. The standard deviation due to counting statistics was generally below 0.5$\%$, 0.3$\%$, and 0.25$\%$ for Ge, Al, and Ce, respectively. Based on the total amounts recorded for Ce, Al, and Ge, any additional In doping was not recorded. WDS results are listed in Table~\ref{tabl2}. Small and nearly identical Ge deficiency is observed among all samples with the In-flux grown sample having Ge concentration closest to 1. However, Al deficiency varies largely among different batches. The first batch listed in Table~\ref{tabl2} shows about 5$\%$ Al deficiency. Crystals grown with Ames Ce show surprisingly large Al deficiency at nearly 20$\%$. In-flux grown crystal is $\sim$10$\%$ Al deficient. There appears to be no correlation between lattice parameters and either Ge or Al deficiency. Both In and Al inclusions were observed in In-grown single crystals.

\begin{figure}[tb]
\centering
\includegraphics[width=1\linewidth]{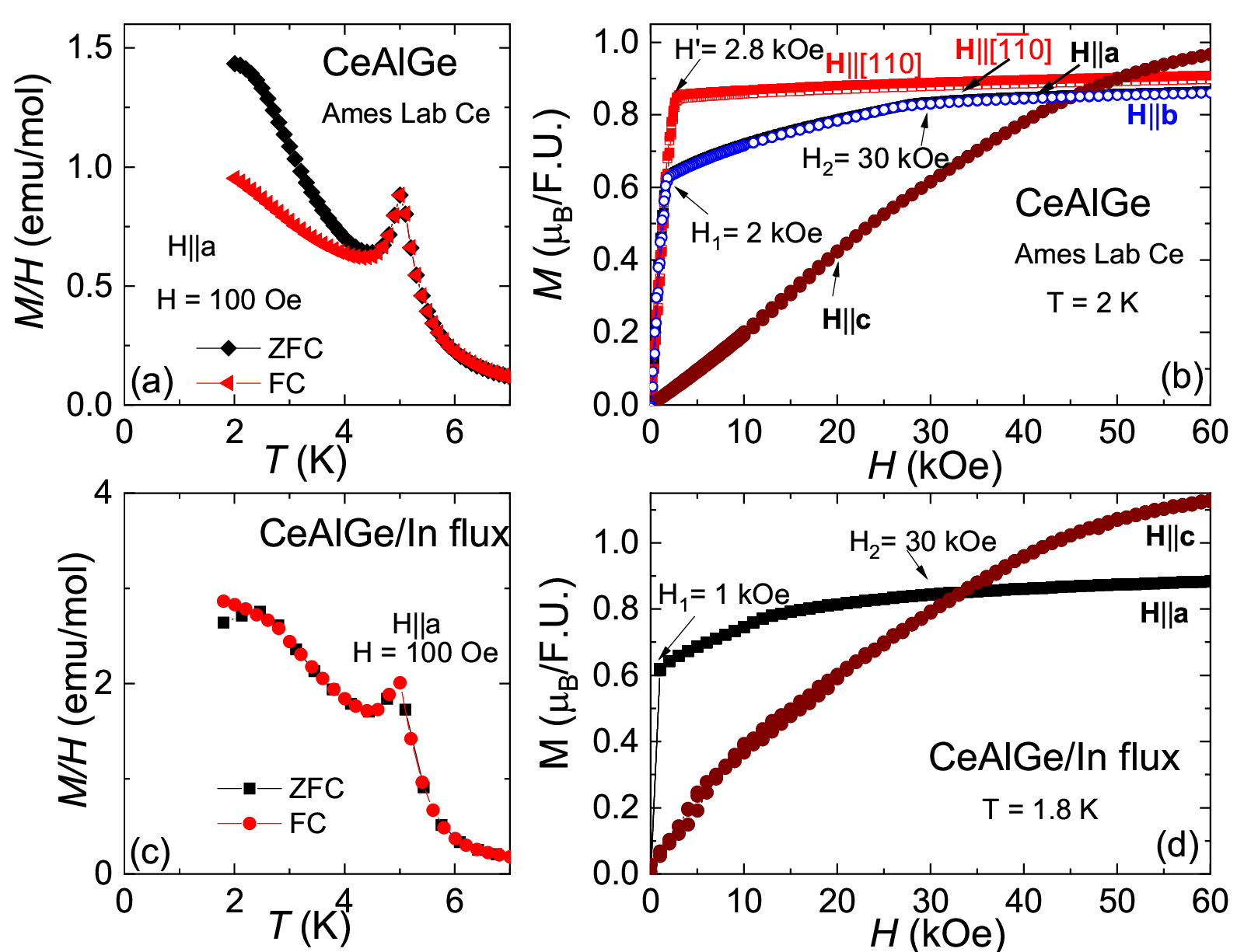}
\caption{\footnotesize (color online) (a) ZFC and FC temperature-dependent magnetization of CeAlGe single crystals grown using Ames cerium. (b) \textit{M(H)} data of CeAlGe/Ames cerium and Al flux, (c) ZFC and FC temperature-dependent magnetization of CeAlGe single crystals grown using In flux, and (d) \textit{M(H)} data of CeAlGe/In flux.}
\label{MHO}
\end{figure}

Zero-field cooled (ZFC) and field-cooled (FC) temperature- and field-dependent magnetization for the batch with Ames Ce/Al-flux and In-flux grown single crystals of CeAlGe are shown in Fig.~\ref{MHO} (a) and (c), respectively. The ordering temperature of 5 K is the same. The effective moments calculated from the Curie-Weiss law fits of the polycrystalline average (not shown here) are consistent with the WDS data. A large difference is seen in the ZFC and FC data below 4.5 K. Temperature-dependent magnetization is also larger at 100 Oe for the In-grown sample. This is consistent with the \textit{M(H)} data shown in Fig.~\ref{MHO}(d) where $H_1$ = 1 kOe (it may even be lower, the data were collected with 1 kOe step) as opposed to $H_1$ = 2 kOe of Ames Ce sample, Fig.~\ref{MHO}(b). $H_1$ critical fields for both samples and $H'$ for the Ames Ce sample are lower than those of the sample presented in the main text. On the other hand, $H_2$ critical fields are somewhat similar. It appears that Al deficiency affects the values of critical fields of spin reorientation $H_1$ and $H'$. 

\textit{M(H)} data for \textbf{H}$\|$\textbf{c} show different behavior. The magnetic moment along the \textit{c}-axis for the In-flux sample is larger than that of the Ames Ce sample and the field at which $M_{H\|c}>M_{H\|a}$ is lower. Two metamagnetic transitions with hysteresis are seen below \textit{H} = 20 kOe for the In-grown sample as well.

\begin{figure}[t]
\centering
\includegraphics[width=1\linewidth]{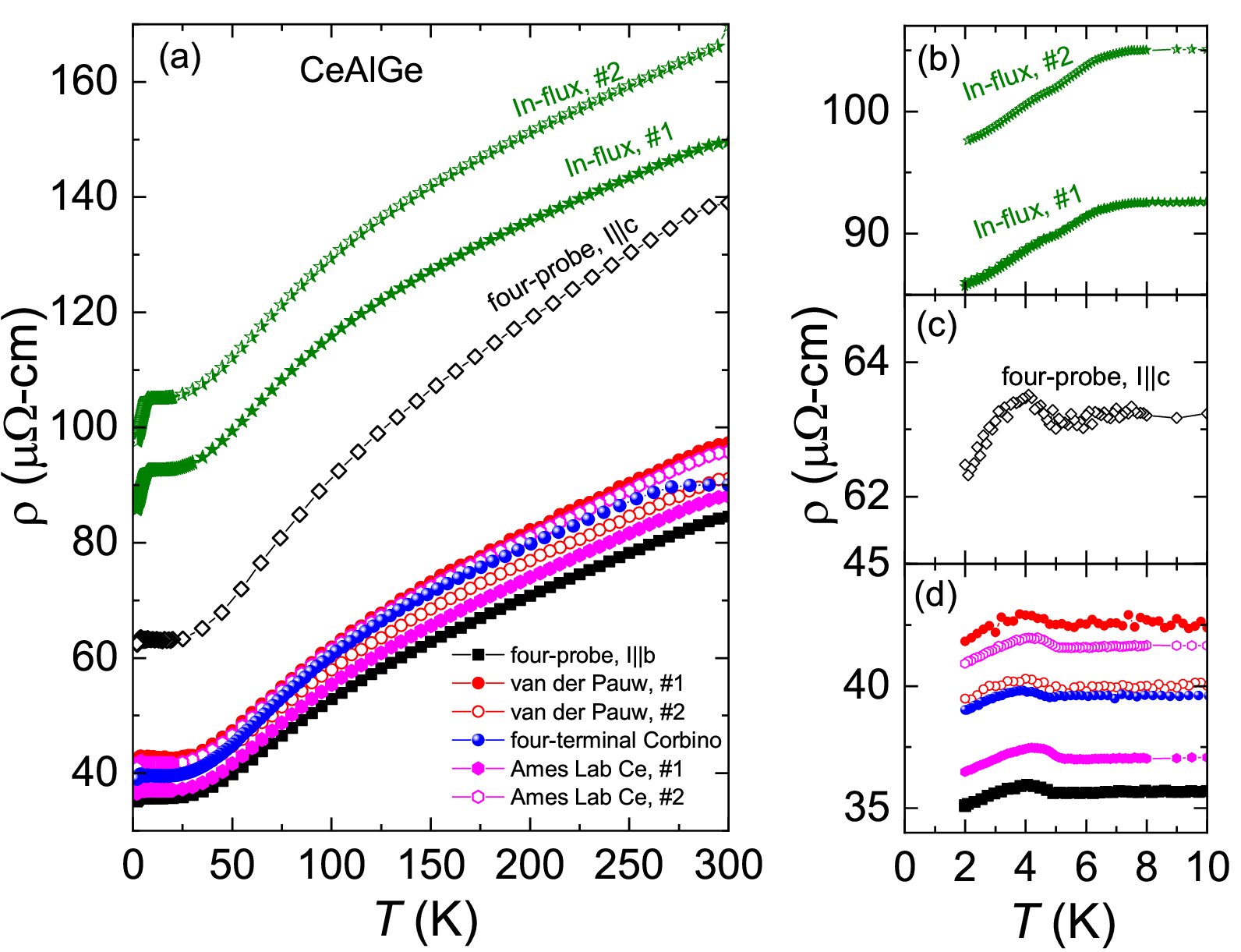}
\caption{\footnotesize (color online) (a) Temperature-dependent resistivity of CeAlGe single crystals grown under different conditions and measured in different electrode configurations. (b), (c), and (d) Low-temperature part of the temperature-dependent resistivity showing the features associated with the magnetic order.}
\label{RT}
\end{figure}

The temperature-dependent resistivity of samples from these batches measured in the standard four-probe configuration are shown in Fig.~\ref{RT} together with those from the batch (measured with different techniques) described in the main text in Fig.~1. The $\rho(T)$ data for Al-flux grown samples show a good agreement. The $\rho(T)$ data for \textbf{I}$\|$\textbf{c} are 1.6 times larger at 300 K and 1.8 times larger at 2 K than that of \textbf{I}$\|$\textbf{b} indicating that the \textit{b} axis is more conductive. In-flux grown samples show larger resistivity values, the feature around 7 K is more pronounced, and the feature associated with the magnetic order is less pronounced compared to those for the Al-flux grown samples. 

\begin{figure}[t]
\centering
\includegraphics[width=1\linewidth]{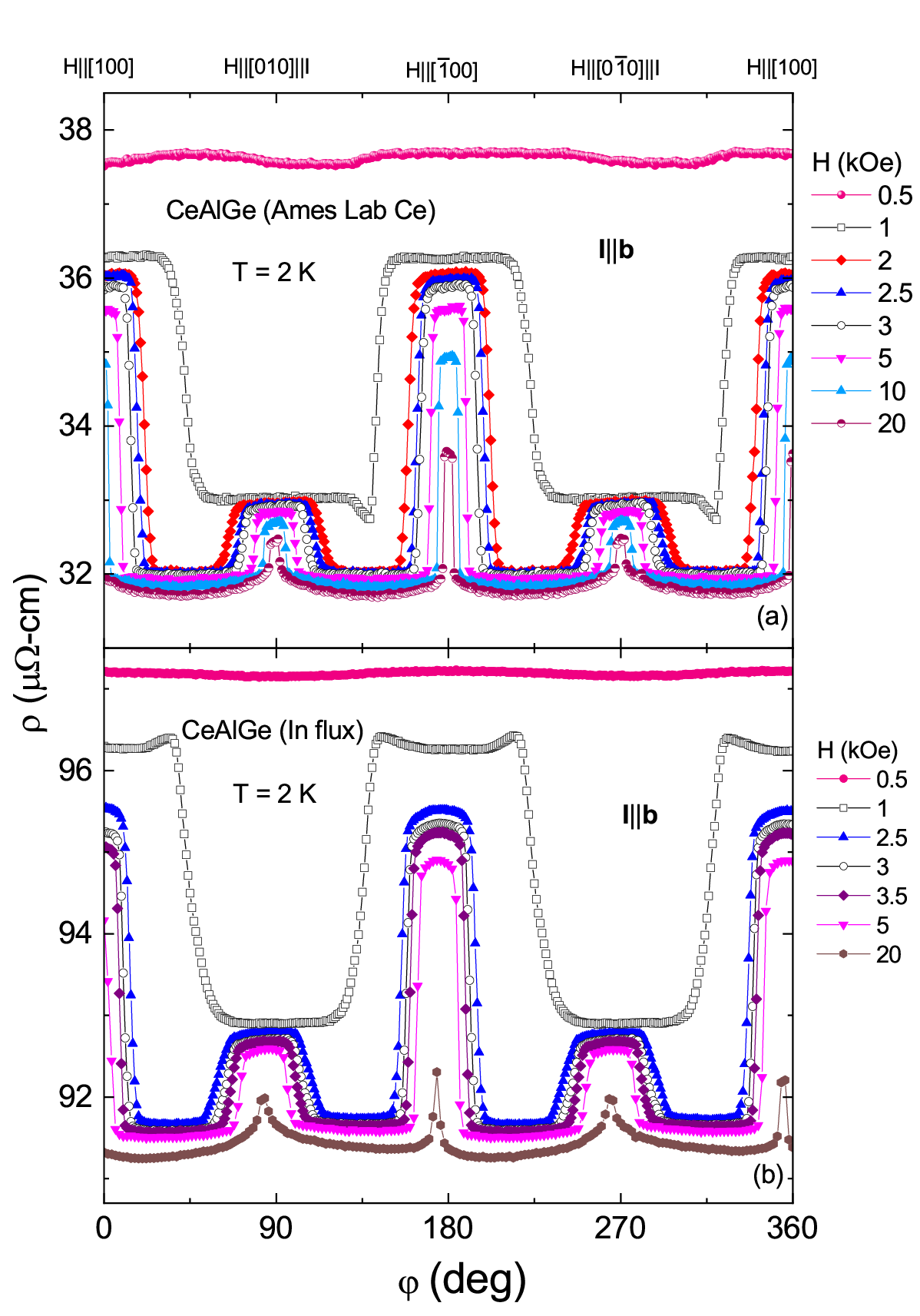}
\caption{\footnotesize (color online) Angle-dependent resistivity of CeAlGe single crystal measured in four-probe wire configuration at 2~K and \textit{H} = const. (a) Ce from Ames Laboratory and Al flux and (b) Ce from Alpha Aesar and In flux were used to grow CeAlGe single crystals.}
\label{DS}
\end{figure}

The AMR data measured in the four-probe configuration with \textbf{I}$\|$\textbf{b} at 2~K at different constant magnetic fields (the field was rotated in the tetragonal \textit{ab} plane) for the Ames Ce/Al-flux and In-flux grown samples are shown in Fig.~\ref{DS}(a) and (b), respectively. The AMR shows similar features for \textit{H}$\geq$ 2 kOe. The same behavior was observed for \textit{H}$\geq$ 5 kOe in the sample discussed in the main text, Fig.~2(b). This corresponds to the region above $H'$ (which is different for these samples). However, the behavior of the AMR data is the same, reflecting the same magnetic state for all three samples above $H'$. On the contrary, the features in the AMR are different below $H'$ for all three samples. For the Ames Ce/Al-flux and In-flux grown samples, the onset of the large and clear two-fold symmetry occurs at a  much lower field of 1 kOe (perhaps even at a smaller field) as opposed to 2.5 kOe for the sample discussed in the main text, due to $H_1$ for these two sample being much lower than that for the sample discussed in the main text. 

Al deficiency appears to affect critical fields $H_1$ and $H'$, making them smaller than that of a stoichiometric sample. This leads to the second regime, common among all samples, of AMR to appear above fields as small as 2 kOe. In addition, Al deficiency seems to affect the spin orientation/structure in the low-field state below $H'$ as is evident by a different functional dependence of the AMR. 

\bibliography{papers}

\end{document}